\documentclass[aps, prc, floatfix, nofootinbib, superscriptaddress, twocolumn]{revtex4-1}

\usepackage{latexsym}
\usepackage{amsmath}
\usepackage{amssymb}
\usepackage{amsfonts}

\usepackage[mathscr,scaled=1.15]{urwchancal}
\DeclareFontFamily{OT1}{pzc}{}
\DeclareFontShape{OT1}{pzc}{m}{it}%
{<-> s * [1.15] pzcmi7t}{}
\DeclareMathAlphabet{\mathpzc}{OT1}{pzc}{m}{it}

\usepackage{color}

\usepackage{supertabular}
\usepackage{placeins}
\usepackage{epsfig}
\usepackage{graphicx}
\usepackage{booktabs}
\usepackage{multirow}

\definecolor{purple}{rgb}{0.5,0,0.5}
\definecolor{blue}{rgb}{0.0,0,0.9}
\definecolor{prdblue}{rgb}{0.133,0.118,0.498}
\usepackage[colorlinks=true, pdfstartview=FitV, linkcolor=prdblue, citecolor= prdblue, urlcolor=prdblue]{hyperref}

\begin{document}

\title{Fully-heavy tetraquarks: $bb\bar{c}\bar{c}$ and $bc\bar{b}\bar{c}$}

\author{Xiaoyun Chen}
\email[]{xychen@jit.edu.cn} \affiliation{Department of Basic
Courses, Jinling Institute of Technology, Nanjing 211169, P. R.
China}

\begin{abstract} \label{abstract}
In the framework of a nonrelativistic chiral quark model, we
continue to study the mass spectra of the fully-heavy
$bb\bar{c}\bar{c}$ and $bc\bar{b}\bar{c}$ tetraquarks. In the
present calculations, two structures, meson-meson
[$\bar{Q}Q$][$\bar{Q}Q$] and diquark-antidiquark
[$QQ$][$\bar{Q}\bar{Q}$] ($Q$ = $c$ or $b$), and their mixing,
along with all possible color, spin configurations are considered.
The calculations suggest that no bound state can be formed for
$bb\bar{c}\bar{c}$ and $bc\bar{b}\bar{c}$ systems. However,
resonances are possible because of the color structure. Several
resonances are predicted and their stabilities are checked using
the real scaling method.
\end{abstract}

\maketitle


\section{Introduction} \label{introduction}
In the past years, experimental searches for the exotic states
beyond the conventional quark model have made great progress. The
observation of $XYZ$ states, such as $X(3872)$~\cite{x3872},
$Y(4260)$~\cite{y4260-1,y4260-2},
$Z_c(3900)$~\cite{zc3900-1,zc3900-2,zc3900-3,zc3900-4},
$Z_b(10610)$~\cite{zb10610}, $P_c^+(4380)$,
$P_c^+(4450)$~\cite{pc} and so on, provided us a good opportunity
to extend our knowledge of the heavy flavor spectroscopy.
Especially those charged quarkonium-like states with heavy flavor
mesons as the decay products, make them the best candidates for
the exotic hadrons.

Recently, the tetraquarks composed of four heavy quarks
$QQ\bar{Q}\bar{Q}$ ($Q$ = $c$ or $b$), have received great
attention. Experimentally, the heavy-flavor states provide some
advantages, because they can be explored with the help of the
efficient triggers such as $J/\psi$. LHCb collaboration is hunting
for the $bb\bar{b}\bar{b}$ tetraquark state and the existence
information needs the further confirmation~\cite{LHCbbbb}. In the
theoretical aspect, there are also many studies on the fully-heavy
tetraquarks. For example, in some
work~\cite{plb773247,epjc78647,prd95034011,arxiv161200012,prd86034004,arxiv180708520,arxiv170607553,arxiv180706040},
it is suggested that there exist stable bound $bb\bar{b}\bar{b}$
and $cc\bar{c}\bar{c}$ states with relatively smaller masses below
the thresholds of the corresponding meson pairs. But some other
work argues to the contrary that there should no bound
$bb\bar{b}\bar{b}$ or $cc\bar{c}\bar{c}$ tetraquark states because
of the lager masses than the thresholds to
decay~\cite{prd252370,prd70014009,prc97035211,prd97094015,prd97054505,epja55106}.
Refs.~\cite{Li:2019uch,Li:2018bkh} also studied the weak decay
properties about tetraquarks $bb\bar{c}\bar{c}$ and
${b\bar{c}}{q\bar{q}}$. Although some of the opinions are quite
different from each other, the researches on the exotic states are
quite important for our understanding the underlying dynamics of
the exotic states and the nature of strong interactions of QCD.

If the $bb\bar{b}\bar{b}$ and $cc\bar{c}\bar{c}$ tetraquark states
do exist in nature, we have strong reason to believe that there
exist more other heavy-flavor tetraquark states. In our previous
work~\cite{epja55106}, we focused on the full-bottom tetraquarks
$bb\bar{b}\bar{b}$ in the framework of the chiral quark model. In
present work, we would like to extend the study to the tetraquarks
$bb\bar{c}\bar{c}$ and $bc\bar{b}\bar{c}$. Although they are still
missing in experiment, the study of the mass spectra of these two
systems will offer information for the further experimental
explorations. Theoretically, in Ref.~\cite{arxiv190102564}, Liu
\emph{et al.} studied the mass spectra of the fully-heavy
tetraquark systems including $bb\bar{c}\bar{c}$ and
$bc\bar{b}\bar{c}$ within a potential model and no bound states
with masses below the corresponding thresholds were found. Recent
studies by Wu \emph{et al.} showed that $bc\bar{b}\bar{c}$ bound
state was found to be possible, but $bb\bar{c}\bar{c}$ state was
not a bound state~\cite{prd97094015}. In Ref.~\cite{prd95054019},
Richard \emph{et al.} also observed that bound $bc\bar{b}\bar{c}$
state might be more favorable than $bb\bar{b}\bar{b}$ and
$cc\bar{c}\bar{c}$. Ref.~\cite{prd86034004} showed that tensor
tetraquark $bc\bar{b}\bar{c}~(2^{++})$ can be observed in both
$B_cB_c$ and $J/\psi \Upsilon(1S)$ modes. It is well-known that
the color magnetic interaction (CMI) of the one-gluon-exchange
plays an important role in the hadron spectrum and hadron-hadron
interactions. Compared with $bb\bar{b}\bar{b}$ tetraquark state,
CMI is beneficial to form compact tetraquarks for
$bc\bar{b}\bar{c}$. Considering the higher thresholds of
$b\bar{c}+b\bar{c}$, $bb\bar{c}\bar{c}$ may also a possible
tetraquark state. Our purpose is firstly to check whether there
are stable bound states in the $bb\bar{c}\bar{c}$ and
$bc\bar{b}\bar{c}$ systems, if not, secondly, we aim to look for
the possible resonances.

In this work, we calculated the mass spectra of the
$bb\bar{c}\bar{c}$ and $bc\bar{b}\bar{c}$ systems in a
nonrelativistic chiral quark model systematically. For
$bb\bar{c}\bar{c}$ state, the possible quantum numbers are
$I(J^P)=0(0^+), 0(1^+)$ and $0(2^+)$. For $bc\bar{b}\bar{c}$
state, it should have definite C-parity, and the allowed quantum
numbers are $I(J^{PC})=0(0^{++}), 0(1^{+-}),0(1^{++})$ and
$0(2^{++})$. For the interaction between the heavy quarks, the
short-distance one-gluon-exchange effects play an important role
now. In the calculations, the meson-meson [$\bar{Q}Q$][$\bar{Q}Q$]
and diquark-antidiquark [$QQ$][$\bar{Q}\bar{Q}$] structures, and
the mixing of them are considered, respectively, along with all
possible color, spin configurations. To distinguish genuine
resonances, we employ the Gaussian expansion method~\cite{GEM}
supplemented by the real scaling method
(stabilization)~\cite{plb633237,prc98045208}. The real scaling
method was often used for analyzing electron-atom and
electron-molecule scattering~\cite{RSM} and was applied in the
quark model calculation
recently~\cite{plb633237,prc98045208,epja55106}.

The paper is organized as follows. In Sec.~\ref{framework}, the chiral quark model and the wave
functions of the four-body system will be introduced briefly. In Sec.~\ref{discussions}, the numerical
results and discussion are presented. A short summary is given in Sec.~\ref{epilogue}.


\section{Quark model and wave functions} \label{framework}
The chiral quark model has been successful both in describing the hadron spectra and hadron-hadron
interactions. The details of the model can be found in Ref.~\cite{094016chen,Vijande:2005}. For
$bb\bar{c}\bar{c}$ and $bc\bar{b}\bar{c}$ full-heavy system, the Hamiltonian of the chiral quark
model consists of three parts: quark rest mass, kinetic energy, and potential energy:
\begin{align}
 H & = \sum_{i=1}^4 m_i  +\frac{p_{12}^2}{2\mu_{12}}+\frac{p_{34}^2}{2\mu_{34}}
  +\frac{p_{1234}^2}{2\mu_{1234}}  \quad  \nonumber \\
  & + \sum_{i<j=1}^4 \left( V_{ij}^{C}+V_{ij}^{G}\right).
\end{align}
The potential energy consists of pieces describing quark confinement (C); one-gluon-exchange (G).
The detailed forms of potentials are shown below (only central parts are presented)
\cite{094016chen}: {\allowdisplaybreaks
\begin{subequations}
\begin{align}
V_{ij}^{C}&= ( -a_c r_{ij}^2-\Delta ) \boldsymbol{\lambda}_i^c
\cdot \boldsymbol{\lambda}_j^c ,  \\
 V_{ij}^{G}&= \frac{\alpha_s}{4} \boldsymbol{\lambda}_i^c \cdot \boldsymbol{\lambda}_{j}^c
\left[\frac{1}{r_{ij}}-\frac{2\pi}{3m_im_j}\boldsymbol{\sigma}_i\cdot
\boldsymbol{\sigma}_j
  \delta(\boldsymbol{r}_{ij})\right],  \\
\delta{(\boldsymbol{r}_{ij})} & =
\frac{e^{-r_{ij}/r_0(\mu_{ij})}}{4\pi r_{ij}r_0^2(\mu_{ij})}.
\end{align}
\end{subequations}}
$m_i$ is the constituent mass of quark/antiquark, and $\mu_{ij}$
is the reduced mass of two interacting quarks and
\begin{equation}
\mu_{1234}=\frac{(m_1+m_2)(m_3+m_4)}{m_1+m_2+m_3+m_4};
\end{equation}
$\mathbf{p}_{ij}=(\mathbf{p}_i-\mathbf{p}_j)/2$,
$\mathbf{p}_{1234}= (\mathbf{p}_{12}-\mathbf{p}_{34})/2$;
$r_0(\mu_{ij}) =s_0/\mu_{ij}$; $\boldsymbol{\sigma}$ are the $SU(2)$ Pauli matrices; $\boldsymbol{\lambda}$,
$\boldsymbol{\lambda}^c$ are $SU(3)$ flavor, color Gell-Mann matrices, respectively; and $\alpha_s$ is an
effective scale-dependent running coupling \cite{Vijande:2005},
\begin{equation}
\alpha_s(\mu_{ij})=\frac{\alpha_0}{\ln\left[(\mu_{ij}^2+\mu_0^2)/\Lambda_0^2\right]}.
\end{equation}
All the parameters are determined by fitting the meson spectrum, from light to heavy; and the resulting
values are listed in Table~\ref{modelparameters}. Table~\ref{mesonmass} gives the masses of some
heavy mesons in the chiral quark model.

\begin{table}[!t]
\begin{center}
\caption{ \label{modelparameters} Model parameters, determined by
fitting the meson spectrum from light to heavy.}
\begin{tabular}{llr}
\hline\hline\noalign{\smallskip}
Quark masses   &$m_u=m_d$    &313  \\
   (MeV)       &$m_s$         &536  \\
               &$m_c$         &1728 \\
               &$m_b$         &5112 \\
\hline
Confinement        &$a_c$ (MeV fm$^{-2}$)         &101 \\
                   &$\Delta$ (MeV)     &-78.3 \\
\hline
OGE                 & $\alpha_0$        &3.67 \\
                   &$\Lambda_0({\rm fm}^{-1})$ &0.033 \\
                  &$\mu_0$(MeV)    &36.98 \\
                   &$s_0$(MeV)    &28.17 \\
\hline\hline
\end{tabular}
\end{center}
\end{table}

\begin{table}[!t]
\begin{center}
\caption{ \label{mesonmass} The masses of some heavy mesons (in
units of MeV). $M_{cal}$ and $M_{exp}$ represents the theoretical
and the experimental masses, respectively.}
\begin{tabular}{ccccccc}
\hline\hline\noalign{\smallskip}
meson     & $\eta_c$  & $J/\psi$ & $\eta_b$ & $\Upsilon$ & $B_c$   & $B_c^*$  \\
\hline
$M_{cal}$ &   2986.3    & 3096.4   & 9334.7   &  9463.9    & 6341.8  & 6395.1        \\
$M_{exp}$ &   2983.6    & 3096.9   & 9399.1    &  9460.3    & 6275.6  &  -       \\
\hline\hline
\end{tabular}
\end{center}
\end{table}

The wave functions of four-quark states for the two structures, diquark-antidiquark and
meson-meson, can be constructed in two steps. For each degree of freedom, first we construct
the wave functions for two-body sub-clusters, then couple the wave functions of two sub-clusters to
obtain the wave functions of four-quark states.

(1) Diquark-antidiquark structure.

For the spin part, the wave functions for two-body sub-clusters are,
\begin{align}
&\chi_{11}=\alpha\alpha,~~
\chi_{10}=\frac{1}{\sqrt{2}}(\alpha\beta+\beta\alpha),~~
\chi_{1-1}=\beta\beta,\nonumber \\
&\chi_{00}=\frac{1}{\sqrt{2}}(\alpha\beta-\beta\alpha),
\end{align}
then the wave functions for four-quark states are obtained,
 {\allowdisplaybreaks
\begin{subequations}\label{spinwavefunctions}
\begin{align}
\chi_{00}^{\sigma
1}&=\chi_{00}\chi_{00},\\
\chi_{00}^{\sigma
2}&=\sqrt{\frac{1}{3}}(\chi_{11}\chi_{1-1}-\chi_{10}\chi_{10}+\chi_{1-1}\chi_{11}),\\
\chi_{11}^{\sigma
3}&=\chi_{00}\chi_{11},\\
 \chi_{11}^{\sigma
4}&=\chi_{11}\chi_{00},\\
\chi_{11}^{\sigma
5}&=\frac{1}{\sqrt{2}}(\chi_{11}\chi_{10}-\chi_{10}\chi_{11}),\\
\chi_{22}^{\sigma 6}&=\chi_{11}\chi_{11},
\end{align}
\end{subequations}}
where the superscript $\sigma i$$(i=1 \sim 6)$ of $\chi$
represents the index of the spin wave functions of four-quark
states. The subscripts of $\chi$ are $SM_S$, the total spin and
the third projection of total spin of the system. $S=0, 1, 2$, and
only one component ($M_S=S$) is shown for a given total spin $S$.

\begin{figure}
\resizebox{0.50\textwidth}{!}{\includegraphics{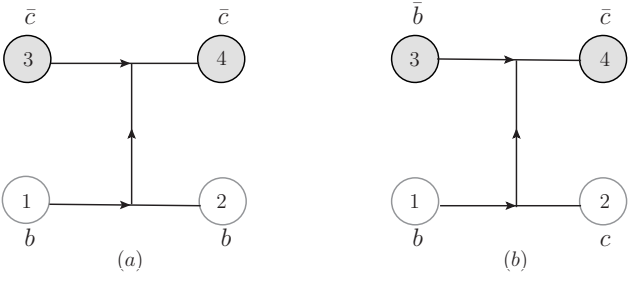}}
\caption{\label{DAstructures} Configurations of the
$bb\bar{c}\bar{c}$ and $bc\bar{b}\bar{c}$ tetraquark states in
pure diquark-antidiquark structure. Figure (a) represents the
configuration of $bb\bar{c}\bar{c}$ state, $[bb][\bar{c}\bar{c}]$;
(b) represents the configuration of $bc\bar{b}\bar{c}$ state,
$[bc][\bar{b}\bar{c}]$.}
\end{figure}

For the flavor part, the configurations of $bb\bar{c}\bar{c}$ and $bc\bar{b}\bar{c}$
states are demonstrated in Fig.~\ref{DAstructures} in diquark-antidiquark structure, and
the wave functions for $bb\bar{c}\bar{c}$ and $bc\bar{b}\bar{c}$ systems take,
\begin{eqnarray}
\chi_{d0}^{f1} & = & (bb)(\bar{c}\bar{c}), \\
\chi_{d0}^{f2} & = & (bc)(\bar{b}\bar{c}),
\end{eqnarray}
respectively. The subscript $d0$ of $\chi$ represents the
diquark-antidiquark structure and isospin ($I=0$).

For the color part, the wave functions of four-quark states must be color singlet $[222]$
and it is obtained as below,
\begin{subequations}
\begin{align}
\chi^{c1}_{d} & =
\frac{\sqrt{3}}{6}(rg\bar{r}\bar{g}-rg\bar{g}\bar{r}+gr\bar{g}\bar{r}-gr\bar{r}\bar{g} \nonumber \\
&~~~+rb\bar{r}\bar{b}-rb\bar{b}\bar{r}+br\bar{b}\bar{r}-br\bar{r}\bar{b} \nonumber \\
&~~~+gb\bar{g}\bar{b}-gb\bar{b}\bar{g}+bg\bar{b}\bar{g}-bg\bar{g}\bar{b}).  \\
\chi^{c2}_{d}&=\frac{\sqrt{6}}{12}(2rr\bar{r}\bar{r}+2gg\bar{g}\bar{g}+2bb\bar{b}\bar{b}
    +rg\bar{r}\bar{g}+rg\bar{g}\bar{r} \nonumber \\
&~~~+gr\bar{g}\bar{r}+gr\bar{r}\bar{g}+rb\bar{r}\bar{b}+rb\bar{b}\bar{r}+br\bar{b}\bar{r} \nonumber \\
&~~~+br\bar{r}\bar{b}+gb\bar{g}\bar{b}+gb\bar{b}\bar{g}+bg\bar{b}\bar{g}+bg\bar{g}\bar{b}).
\end{align}
\end{subequations}
Where, $\chi_{d}^{c1}$ and $\chi_{d}^{c2}$ represents the color antitriplet-triplet
($\bar{3}\times3$) and sextet-antisextet ($6\times\bar{6}$) coupling, respectively. The detailed
coupling process for the color wave functions can refer to our previous work \cite{054022chen}.

(2) Meson-meson structure.

For the spin part, the wave functions are the same as those of the diquark-antidiquark structure,
Eq.~(\ref{spinwavefunctions}).

\begin{figure}
\resizebox{0.50\textwidth}{!}{\includegraphics{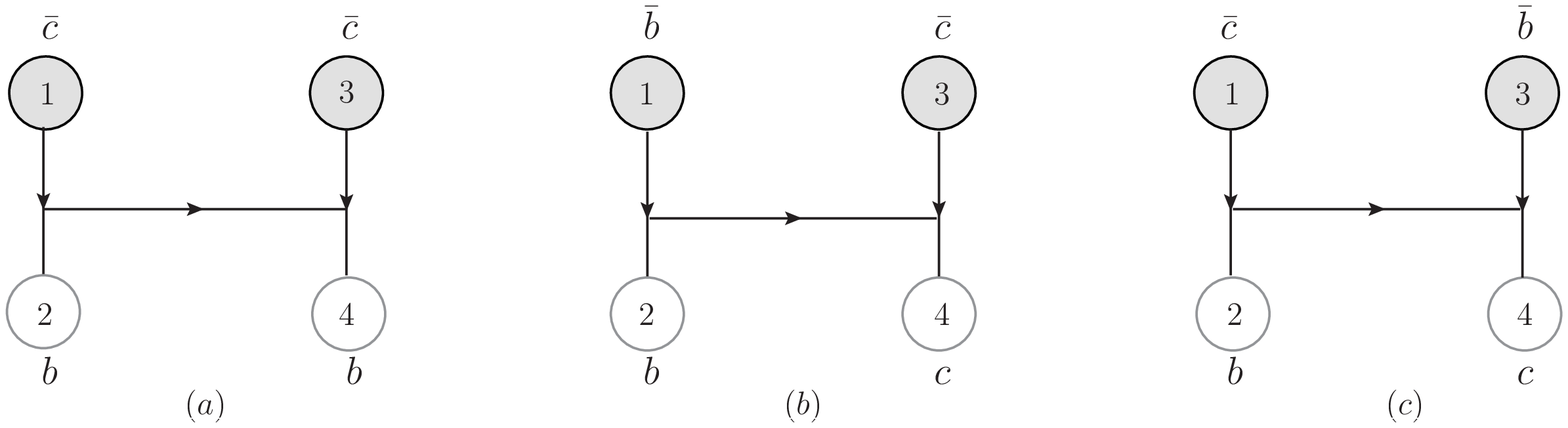}}
\caption{\label{MMstructures} Three configurations of the $bb\bar{c}\bar{c}$ and
$bc\bar{b}\bar{c}$ tetraquark states in meson-meson structure. (a) the only configuration of
$bb\bar{c}\bar{c}$ system $(\bar{c}b)(\bar{c}b)$; (b) and (c) two configurations of
$bc\bar{b}\bar{c}$ system, $(\bar{b}b)(\bar{c}c)$ and $(\bar{c}b)(\bar{b}c)$, respectively.}
\end{figure}

For the flavor part, there are three wave functions, one function for $bb\bar{c}\bar{c}$ system,
\begin{equation}
\chi_{m0}^{f1}=(\bar{c}b)(\bar{c}b),
\end{equation}
and two functions for $bc\bar{b}\bar{c}$ system,
\begin{eqnarray}
\chi_{m0}^{f2} & = & (\bar{b}b)(\bar{c}c), \\
\chi_{m0}^{f3} & = & (\bar{c}b)(\bar{b}c).
\end{eqnarray}
The subscript $m0$ of $\chi$ represents the meson-meson structure
and isospin equals zero. Fig.~\ref{MMstructures} shows the
meson-meson structure of $bb\bar{c}\bar{c}$ and $bc\bar{b}\bar{c}$
systems.

For the color part, the wave functions of four-quark states in the meson-meson structure are,
\begin{subequations}
\begin{align}
\chi_{m}^{c1}&=\frac{1}{3}(\bar{r}r+\bar{g}g+\bar{b}b)(\bar{r}r+\bar{g}g+\bar{b}b),\\
\chi_{m}^{c2}&=\frac{\sqrt{2}}{12}(3\bar{b}r\bar{r}b+3\bar{g}r\bar{r}g+3\bar{b}g\bar{g}b+3\bar{g}b\bar{b}g+3\bar{r}g\bar{g}r \nonumber \\
&~~~+3\bar{r}b\bar{b}r+2\bar{r}r\bar{r}r+2\bar{g}g\bar{g}g+2\bar{b}b\bar{b}b-\bar{r}r\bar{g}g \nonumber\\
&~~~-\bar{g}g\bar{r}r-\bar{b}b\bar{g}g-\bar{b}b\bar{r}r-\bar{g}g\bar{b}b-\bar{r}r\bar{b}b).
\end{align}
\end{subequations}
Where, $\chi_{m}^{c1}$ and $\chi_{m}^{c2}$ represents the color singlet-singlet ($1\times1$) and
color octet-octet ($8\times8$) coupling, respectively. The details refer to our previous work
\cite{054022chen}.

As for the orbital wave functions, they can be constructed by coupling the orbital wave function
for each relative motion of the system,
\begin{equation}\label{spatialwavefunctions}
\Psi_{L}^{M_{L}}=\left[[\Psi_{l_1}({\bf r}_{12})\Psi_{l_2}({\bf
r}_{34})]_{l_{12}}\Psi_{L_r}({\bf r}_{1234}) \right]_{L}^{M_{L}},
\end{equation}
where $l_1$ and $l_2$ is the angular momentum of two sub-clusters, respectively.
$\Psi_{L_r}(\mathbf{r}_{1234})$ is the wave function of the relative motion between two
sub-clusters with orbital angular momentum $L_r$. $L$ is the total orbital angular momentum
of four-quark states. Here for the low-lying $bb\bar{c}\bar{c}$ and $bc\bar{b}\bar{c}$ state,
all angular momentum ($l_1, l_2, L_r, L$) are taken as zero. The used Jacobi coordinates
are defined as,
\begin{align}\label{jacobi}
{\bf r}_{12}&={\bf r}_1-{\bf r}_2, \nonumber \\
{\bf r}_{34}&={\bf r}_3-{\bf r}_4, \nonumber\\
{\bf r}_{1234}&=\frac{m_1{\bf r}_1+m_2{\bf
r}_2}{m_1+m_2}-\frac{m_3{\bf r}_3+m_4{\bf r}_4}{m_3+m_4}.
\end{align}
For diquark-antidiquark structure, the quarks are numbered as $1, 2$, and the antiquarks are
numbered as $3, 4$; for meson-meson structure, the antiquark and quark in one cluster are
marked as $1, 2$, the other antiquark and quark are marked as $3, 4$. In the two structure
coupling calculation, the indices of quarks, antiquarks in diquark-antidiquark structure will be
changed to be consistent with the numbering scheme in meson-meson structure.
In GEM, the spatial wave function is expanded by Gaussians~\cite{GEM}:
\begin{subequations}
\label{radialpart}
\begin{align}
\Psi_{l}^{m}(\mathbf{r}) & = \sum_{n=1}^{n_{\rm max}} c_{n}\psi^G_{nlm}(\mathbf{r}),\\
\psi^G_{nlm}(\mathbf{r}) & = N_{nl}r^{l}
e^{-\nu_{n}r^2}Y_{lm}(\hat{\mathbf{r}}),
\end{align}
\end{subequations}
where $N_{nl}$ are normalization constants,
\begin{align}
N_{nl}=\left[\frac{2^{l+2}(2\nu_{n})^{l+\frac{3}{2}}}{\sqrt{\pi}(2l+1)}
\right]^\frac{1}{2}.
\end{align}
$c_n$ are the variational parameters, which are determined dynamically. The Gaussian size
parameters are chosen according to the following geometric progression
\begin{equation}\label{gaussiansize}
\nu_{n}=\frac{1}{r^2_n}, \quad r_n=r_1a^{n-1}, \quad
a=\left(\frac{r_{n_{\rm max}}}{r_1}\right)^{\frac{1}{n_{\rm
max}-1}}.
\end{equation}
This procedure enables optimization of the expansion using just a small numbers of Gaussians.
Finally, the complete channel wave function for the four-quark system for diquark-antidiquark
structure is written as
\begin{align}\label{diquarkpsi}
&\Psi_{IJ,i,j,k}^{M_IM_J}={\cal
A}_1[\Psi_{L}^{M_{L}}\chi_{SM_{S}}^{\sigma
i}]_{J}^{M_J}\chi_{d0}^{fj}\chi^{ck}_{d},
  \nonumber \\
&(i=1\sim6; j=1,2; k=1,2; S=0,1,2),
\end{align}
where ${\cal A}_1$ is the antisymmetrization operator, for
$bb\bar{c}\bar{c}$ system,
\begin{equation}
{\cal A}_1=\frac{1}{2}(1-P_{12}-P_{34}+P_{12}P_{34}).
\end{equation}

For meson-meson structure, the complete wave function is written as
\begin{align}
&\Psi_{IJ,i,j,k}^{M_IM_J}= {\cal
A}_2[\Psi_{L}^{M_{L}}\chi_{SM_{S}}^{\sigma
i}]_{J}^{M_J}\chi_{m0}^{fj}\chi^{ck}_{m},\nonumber \label{mesonpsi}\\
&(i=1\sim6; j=1,2,3; k=1,2; S=0,1,2),
\end{align}
where ${\cal A}_2$ is the antisymmetrization operator, for
$bb\bar{c}\bar{c}$ system,
\begin{equation}
{\cal A}_2=\frac{1}{2}(1-P_{13}-P_{24}+P_{13}P_{24}).
\end{equation}

Lastly, the eigenenergies of the four-quark system are obtained by solving a
Schr\"{o}dinger equation:
\begin{equation}
    H \, \Psi^{\,M_IM_J}_{IJ}=E^{IJ} \Psi^{\,M_IM_J}_{IJ},
\end{equation}
where $\Psi^{\,M_IM_J}_{IJ}$ is the wave function of the four-quark states, which is the
linear combinations of the above channel wave functions, Eq.~(\ref{diquarkpsi}) in the
diquark-anti-diquark structure or Eq.~(\ref{mesonpsi}) in the meson-meson structure, or
both wave functions of Eq.~(\ref{diquarkpsi}) and (\ref{mesonpsi}), respectively.

\section{Results and discussions} \label{discussions}
In the present work, we calculated the mass spectra of the
$bb\bar{c}\bar{c}$ and $bc\bar{b}\bar{c}$ systems with allowed
quantum numbers in the nonrelativistic quark model. Two structures
of four-quark states, meson-meson and diquark-antidiquark, and the
mixing of them are investigated, respectively. All possible color,
and spin configurations are also considered. For example, for
meson-meson structure, two color configurations, color
singlet-singlet ($1 \times 1$) and octet-octet ($8 \times 8$) are
employed; for diquark-antidiquark structure, color
antitriplet-triplet ($\bar{3} \times 3$) and sextet-antisextet ($6
\times \bar{6}$) are taken into account. For $bb\bar{c}\bar{c}$
state, the wave functions need to be antisymmetrized. All the
allowed channels are demonstrated in Table~\ref{WFunctions}. For
$bc\bar{b}\bar{c}$, there is no need to consider the
antisymmetrization because of no identical quarks. Because the
hamiltonian of the system is invariant under the charge conjugate,
the $C$-parity is a good quantum number and is shown in the table
for $bc\bar{b}\bar{c}$ system. All possible channels are also
showed in Table~\ref{WFunctions}.
\begin{table}[!t]
\begin{center}
\caption{ \label{WFunctions} The allowed channels of $bb\bar{c}\bar{c}$ and $bc\bar{b}\bar{c}$
systems in meson-meson (M-M) and diquark-antidiquark (D-A) structures.
$\chi^{\sigma 34 f3 c1,2\mp}_{m}=\sqrt{\frac{1}{2}}(\chi^{\sigma 3}_{1}\chi^{f3}_{m0}\chi^{c1,2}_{m}
 \pm \chi^{\sigma 4}_{1}\chi^{f3,C}_{m0}\chi^{c1,2,C}_{m})$, $\chi^{\sigma 34 f2 c1,2\mp}_{d}=\sqrt{\frac{1}{2}}(\chi^{\sigma 3}_{1}\chi^{f2}_{d0}\chi^{c1,2}_{d}
 \pm \chi^{\sigma 4}_{1}\chi^{f2,C}_{d0}\chi^{c1,2,C}_{d})$. The wave functions with superscript ``C",
 for example $\chi^{f3,C}_{m0}$, are the charge conjugate of the corresponding wave function without superscript ``C".}
\begin{tabular}{ccccc} \hline \hline
   \multicolumn{4}{c}{$bb\bar{c}\bar{c}$} \\ \hline
   $I(J^P)$ & $0(0^+)$ & $0(1^+)$ & $0(2^+)$ & \\ \hline
   M-M  & $\chi^{\sigma 1,2}_{0}\chi^{f1}_{m0}\chi^{c1,2}_{m}$  & $\chi^{\sigma 3,4}_{1}\chi^{f1}_{m0}\chi^{c1,2}_{m}$
              & $\chi^{\sigma 6}_{2}\chi^{f1}_{m0}\chi^{c1,2}_{m}$    \\
   D-A  & $\chi^{\sigma 1}_{0}\chi^{f1}_{d0}\chi^{c2}_{d}$      & $\chi^{\sigma 5}_{1}\chi^{f1}_{d0}\chi^{c1}_{d}$
              & $\chi^{\sigma 6}_{2}\chi^{f1}_{d0}\chi^{c1}_{d}$    \\
        & $\chi^{\sigma 2}_{0}\chi^{f1}_{d0}\chi^{c1}_{d}$      &   &     \\ \hline
   \multicolumn{5}{c}{$bc\bar{b}\bar{c}$} \\ \hline
   $I(J^{PC})$ & $0(0^{++})$ & $0(1^{+-})$ & $0(1^{++})$ & $0(2^{++})$ \\ \hline
   M-M  & $\chi^{\sigma 1,2}_{0}\chi^{f2,3}_{m0}\chi^{c1,2}_{m}$    & $\chi^{\sigma 3,4}_{1}\chi^{f2}_{m0}\chi^{c1,2}_{m}$
              & $\chi^{\sigma 5}_{1}\chi^{f2}_{m0}\chi^{c1,2}_{m}$  & $\chi^{\sigma 6}_{2}\chi^{f2,3}_{m0}\chi^{c1,2}_{m}$ \\
        &     & $\chi^{\sigma 34 f3 c1,2-}_{m}$  & $\chi^{\sigma 34 f3 c1,2+}_{m}$  &    \\
        &     & $\chi^{\sigma 5}_{1}\chi^{f3}_{m0}\chi^{c1,2}_{m}$ &   &    \\
   D-A  & $\chi^{\sigma 1,2}_{0}\chi^{f2}_{d0}\chi^{c1,2}_{d}$      & $\chi^{\sigma 34 f2 c1,2-}_{d}$
              & $\chi^{\sigma 34 f2 c1,2+}_{d}$  & $\chi^{\sigma 6}_{2}\chi^{f2}_{d0}\chi^{c1,2}_{d}$  \\
        &     & $\chi^{\sigma 5}_{1}\chi^{f2}_{d0}\chi^{c1,2}_{d}$  &    &    \\
   \hline \hline
\end{tabular}
\end{center}
\end{table}

\begin{table}[!t]
\begin{center}
\caption{ \label{meCS} The matrix elements of color and spin operators.
 $O_{ij}=\boldsymbol{\lambda}_i\cdot \boldsymbol{\lambda}_j$ for color, and
 $O_{ij}=\boldsymbol{\sigma}_i\cdot \boldsymbol{\sigma}_j$ for spin.}
\begin{tabular}{ccccccccc} \hline \hline
  & \multicolumn{4}{c}{color} & \multicolumn{4}{c}{spin} \\ \hline
  & \multicolumn{2}{c}{D-A} & \multicolumn{2}{c}{M-M} & \multicolumn{2}{c}{D-A} & \multicolumn{2}{c}{M-M}
  \\ \hline
  & $\bar{3}\otimes 3$~ & $6\otimes \bar{6}$~ & $1\otimes 1$~ &  $8\otimes 8$~
  & $0\otimes 0$~ &   $1\otimes 1$~ & $0\otimes 0$~ &  $1\otimes 1$  \\ \hline
$\langle O_{12}\rangle$  & $-\frac{4}{3}$ & $-\frac{10}{3}$ & $-\frac{16}{3}$ & $\frac{2}{3}$
  & $0$ &  $-2$ & $-3$ & $1$ \\
$\langle O_{13}\rangle$  & $-\frac{8}{3}$ & $\frac{4}{3}$ & $0$ & $-\frac{4}{3}$
  & $-3$ &  $1$ & $0$ & $-2$ \\
$\langle O_{14}\rangle$  & $-\frac{4}{3}$ & $-\frac{10}{3}$ & $0$ & $-\frac{14}{3}$
  & $0$ &  $-2$ & $0$ & $-2$ \\
$\langle O_{23}\rangle$  & $-\frac{4}{3}$ & $-\frac{10}{3}$ & $0$ & $-\frac{14}{3}$
  & $0$ &  $-2$ & $0$ & $-2$ \\
$\langle O_{24}\rangle$  & $-\frac{8}{3}$ & $\frac{4}{3}$ & $0$ & $-\frac{4}{3}$
  & $-3$ &  $1$ & $0$ & $-2$ \\
$\langle O_{34}\rangle$  & $-\frac{4}{3}$ & $-\frac{10}{3}$ & $-\frac{16}{3}$ & $\frac{2}{3}$
  & $0$ &  $-2$ & $-3$ & $1$ \\
  \hline \hline
\end{tabular}
\end{center}
\end{table}

\begin{table*}[!t]
\begin{center}
\caption{ \label{CMI} The matrix elements of CMI in the unit of
$1/m_b^2$, $x=m_b/m_c$. $\Delta_{CMI}$ is the difference of the
matrix elements between tetraquark system and two-meson pairs. The
expressions with underline are the thresholds of the corresponding
systems.}
\begin{tabular}{ccccccccc} \hline \hline
 color &  \multicolumn{2}{c}{$\bar{3}\otimes 3$} & \multicolumn{2}{c}{$6\otimes \bar{6}$}
  & \multicolumn{2}{c}{$1\otimes 1$} & \multicolumn{2}{c}{$8\otimes 8$}  \\ \hline
 spin & $0\otimes 0$ &  $1\otimes 1$ &  $0\otimes 0$ &   $1\otimes 1$ &
   $0\otimes 0$ &  $1\otimes 1$ &  $0\otimes 0$ &   $1\otimes 1$ \\ \hline
 $bb\bar{c}\bar{c}$ & $-8-8x^2$ & $-\frac{32x}{3}+\frac{8}{3}+\frac{8x^2}{3}$
  & $4+4x^2$ & $-\frac{80x}{3}+\frac{4}{3}+\frac{4x^2}{3}$
  & $-32x$ &  $\frac{32x}{3}$ & $4x$ & $-\frac{60x}{3}-\frac{8}{3}-\frac{8x^2}{3}$ \\
 $b\bar{c}+b\bar{c}$ & & & & & \underline{$-32x$} & $\frac{32x}{3}$ & & \\
 $\Delta_{CMI}$ & $16x-8(x-1)^2$ & $\frac{64x}{3}+\frac{8}{3}+\frac{8x^2}{3}$ & $4+4x^2+32x$
  & $\frac{16x}{3}+\frac{4}{3}+\frac{4x^2}{3}$ & 0 & $\frac{128x}{3}$ & $36x$ & $\frac{20x}{3}-\frac{8}{3}(x-1)^2$  \\
   \hline
 $bc\bar{b}\bar{c}$ & $-16x$ & $-\frac{8}{3}-\frac{8x^2}{3}$ &
  $4x$ & $-\frac{32x}{3}-\frac{20}{3}-\frac{20x^2}{3}$
  & $-16-16x^2$ & $\frac{16}{3}+\frac{16x^2}{3}$  & $2+2x^2$ & $-\frac{72x}{3}-\frac{2}{3}-\frac{2x^2}{3}$ \\
  $b\bar{c}+c\bar{b}$ & & & & & $-32x$ & $\frac{32x}{3}$ & & \\
 $b\bar{b}+c\bar{c}$ & & & & & \underline{$-16-16x^2$} & $\frac{16}{3}+\frac{16x^2}{3}$ & & \\
 $\Delta_{CMI}$ & $16+16x^2-16x$ & $\frac{40}{3}+\frac{40x^2}{3}$ & $16+16x^2+4x$
  & $\frac{28}{3}+\frac{28x^2}{3}-\frac{32x}{3}$ & 0 & $\frac{64}{3}+\frac{64x^2}{3}$ & $18+18x^2$
  & $\frac{46}{3}+\frac{46x^2}{3}-\frac{72x}{3}$ \\
  \hline \hline
\end{tabular}
\end{center}
\end{table*}

For full-heavy flavor system, the large masses of $b$ and $c$-quark prevent the appearance of the
Goldstone boson exchanges, only gluon exchanges are included. It is helpful to understand the numerical
results by analyzing qualitatively the properties of the interactions between quarks in the system.
Table~\ref{meCS} gives the matrix elements of color operators and spin operators (only the results for
total spin S=0 are given here). With the help of these matrix elements, we can estimate roughly
the binding energy of the system. For single meson, the matrix element of color operator is
\begin{eqnarray}
 & & \langle \boldsymbol{\lambda}_1 \cdot \boldsymbol{\lambda}_2 \rangle
 =-16/3.
\end{eqnarray}
For tetraquark states, the matrix element of color operator is
\begin{eqnarray}
 & & \sum_{i>j=1}^{4} \langle \boldsymbol{\lambda}_i \cdot \boldsymbol{\lambda}_j \rangle =-32/3,
 ~~~~\mbox{for D-A and M-M.}
\end{eqnarray}
So the color matrix elements are exactly same for  the tetraquark
system and two-meson pairs and the pure color interaction cannot
contribute the binding energy of the tetraquark system.

For CMI (color magnetic interaction), all the matrix elements are
given in the Table~\ref{CMI}. From the table, we can see that the
difference of the CMI matrix elements between tetraquark system
and two-meson pairs are not smaller than 0, so CMI cannot lead to
deep bound state. It is worth to note that $\Delta_{CMI}$ for
color-spin configuration $\bar{3}\otimes 3$, $0\otimes 0$ and
$8\otimes 8$, $1\otimes 1$ may be negative if $x$ is large enough,
which means that the bound states are more possible in
$QQ\bar{q}\bar{q}$ systems. Some previous work, for example
Ref.~\cite{ycyang}, obtained several bound states in these
systems.

The numerical results of
$bb\bar{c}\bar{c}$ and $bc\bar{b}\bar{c}$ systems are shown in Tables~\ref{results1} and \ref{results2},
respectively. $E_{cc}$ represents the ground state energy for each state after considering the all possible
color and spin channels (refer to Table~\ref{WFunctions}). For $bc\bar{b}\bar{c}$ system,
the states with different C-parity are separated.

\begin{table}[!t]
\begin{center}
\caption{ \label{results1} The results of $bb\bar{c}\bar{c}$ state
in pure meson-meson structure, diquark-antidiquark structure, and
in considering the mixing of two structures, respectively.
"$E_{th}^{theo}$" represents the theoretical thresholds. The
masses are all in units of MeV.} \setlength{\tabcolsep}{0.5mm}{
\begin{tabular}{ccccc} \hline \hline
Structure &$J^P$  &$E_{cc}$  &thresholds &$E_{th}^{theo}$ \\
\hline
$[\bar{c}b][\bar{c}b]$ &$0^+$  &12683.9   &$2B_c^-$              &12683.6    \\
                       &$1^+$  &12737.4   &$B_c^-B_c^{*-}$       &12736.9      \\
                       &$2^+$  &12790.7   &$2B_c^{*-}$           &12790.2     \\
$[bb][\bar{c}\bar{c}]$ &$0^+$  &12891.5   &$2B_c^-$              &12683.6   \\
                       &$1^+$  &12897.6   &$B_c^-B_c^{*-}$       &12736.9 \\
                       &$2^+$  &12904.5   &$2B_c^{*-}$           &12790.2 \\
$[\bar{c}b][\bar{c}b] \otimes [bb][\bar{c}\bar{c}]$
                       &$0^+$  &12683.9   &$2B_c^-$               &12683.6     \\
                       &$1^+$  &12737.4   &$B_c^-B_c^{*-}$        &12736.9   \\
                       &$2^+$  &12790.7   &$2B_c^{*-}$            &12790.2    \\
\hline \hline
\end{tabular}}
\end{center}
\end{table}

\begin{table}[!t]
\begin{center}
\caption{ \label{results2} The results of $bc\bar{b}\bar{c}$ state
in pure meson-meson structure, diquark-antidiquark structure, and
in considering the mixing of two structures, respectively.
"$E_{th}^{theo}$" represents the theoretical thresholds. The
masses are all in units of MeV.} \setlength{\tabcolsep}{0.5mm}{
\begin{tabular}{ccccc} \hline \hline
Structure &$J^{PC}$  &$E_{cc}$  &thresholds &$E_{th}^{theo}$ \\
\hline
$[\bar{b}b][\bar{c}c]$ &$0^{++}$  &12321.5  &$\eta_b\eta_c$              &12321.0    \\
                       &$1^{++}$  &12561.0  &$\Upsilon J/\psi$           &12560.3    \\
                       &$1^{+-}$  &12431.6  &$\eta_bJ/\psi$              &12431.1                            \\
                       &$2^{++}$  &12560.9  &$\Upsilon J/\psi$           &12560.3                            \\
$[\bar{c}b][\bar{b}c]$ &$0^{++}$  &12684.0  &$B_c^+B_c^-$                &12683.6   \\
                       &$1^{++}$  &12737.3  &$B_c^+B_c^{*-}$             &12736.9                         \\
                       &$1^{+-}$  &12737.3  &$B_c^+B_c^{*-}$             &12736.9                         \\
                       &$2^{++}$  &12790.6  &$B_c^{*+}B_c^{*-}$          &12790.2                         \\
$[bc][\bar{b}\bar{c}]$ &$0^{++}$  &12746.2  &$\eta_b\eta_c$              &12321.0 \\
                       &$1^{++}$  &12804.2  &$\Upsilon J/\psi$           &12560.3                        \\
                       &$1^{+-}$  &12776.4  &$\eta_bJ/\psi$              &12431.1                        \\
                       &$2^{++}$  &12809.3  &$\Upsilon J/\psi$           &12560.3                        \\
$[\bar{b}b][\bar{c}c] \otimes [\bar{c}b][\bar{b}c] \otimes
[bc][\bar{b}\bar{c}]$  &$0^{++}$  &12321.5  &$\eta_b\eta_c$       &12321.0\\
                       &$1^{++}$  &12561.0  &$\Upsilon J/\psi$    &12560.3                        \\
                       &$1^{+-}$  &12431.6  &$\eta_bJ/\psi$       &12431.1                        \\
                       &$2^{++}$  &12561.0  &$\Upsilon J/\psi$    &12560.3                       \\
\hline \hline
\end{tabular}}
\end{center}
\end{table}

From the Table~\ref{results1}, we found that the lowest energies
of $0^+$, $1^+$ and $2^+$ in the meson-meson structure are a
little higher than the relevant thresholds. In the
diquark-antidiquark structure, the energies are all much lager
than those in the meson-meson structure. The effects of the
two-structure mixing seem to be tiny. So we cannot find the bound
states of $bb\bar{c}\bar{c}$ tetraquark in the present
calculation. For $bc\bar{b}\bar{c}$ system, with the lower
threshold $b\bar{b}+c\bar{c}$ compared with $b\bar{c}+b\bar{c}$,
it may be much harder to form a bound state. In
Table~\ref{results2}, the lowest energies of the three structures
of $bc\bar{b}\bar{c}$ system are all larger than the corresponding
thresholds. Situations are not changed in considering the mixing
of the three quark structures. No bound states are found and
recent study by Liu \emph{et al.}~\cite{arxiv190102564} also draws
the same conclusion with ours.

\begin{figure*}
\centering
    \begin{minipage}[c]{0.45\textwidth}
    \centering
    \includegraphics[height=6.0cm,width=8.5cm]{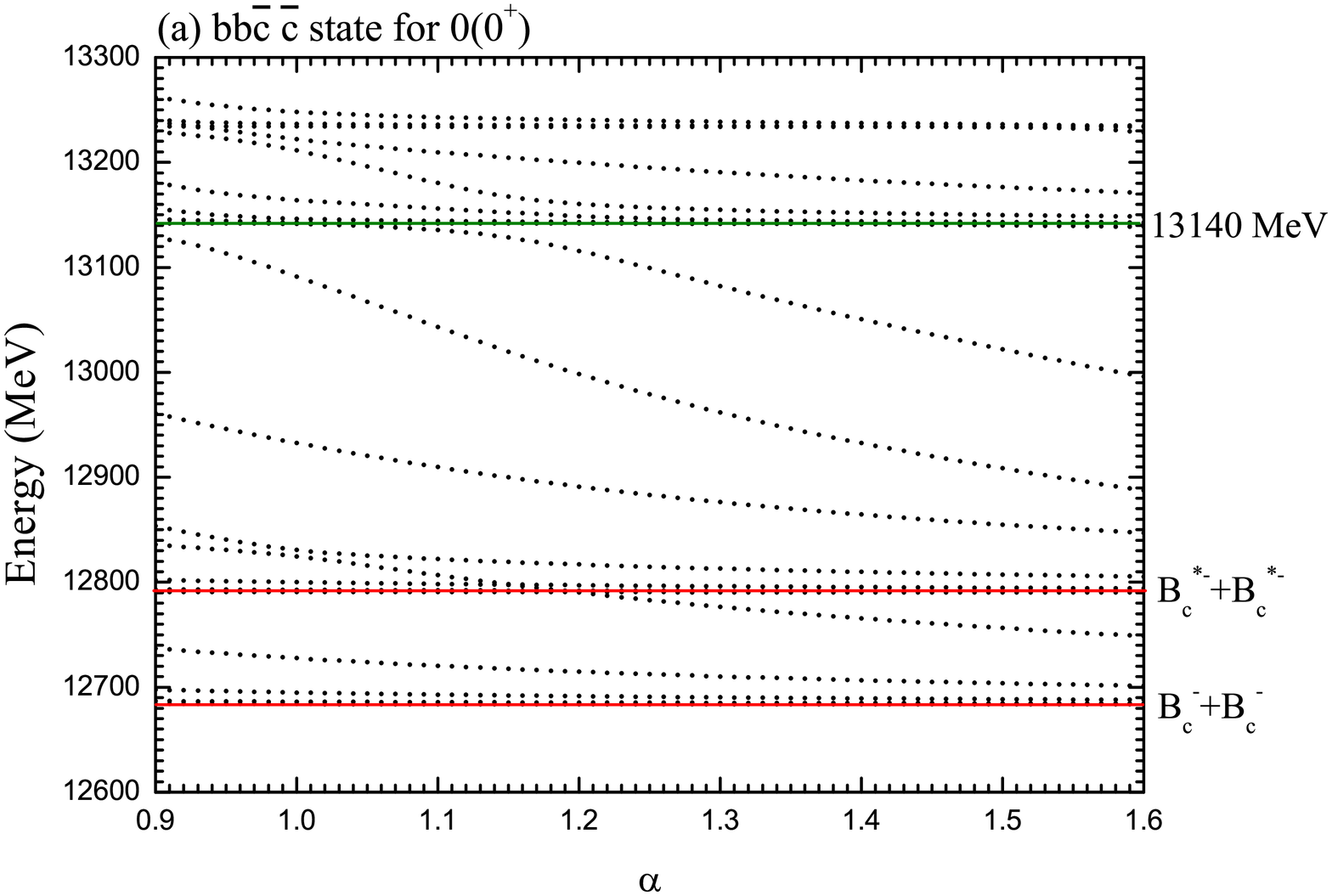}
    \end{minipage}
    \begin{minipage}[c]{0.45\textwidth}
    \centering
    \includegraphics[height=6.0cm,width=8.5cm]{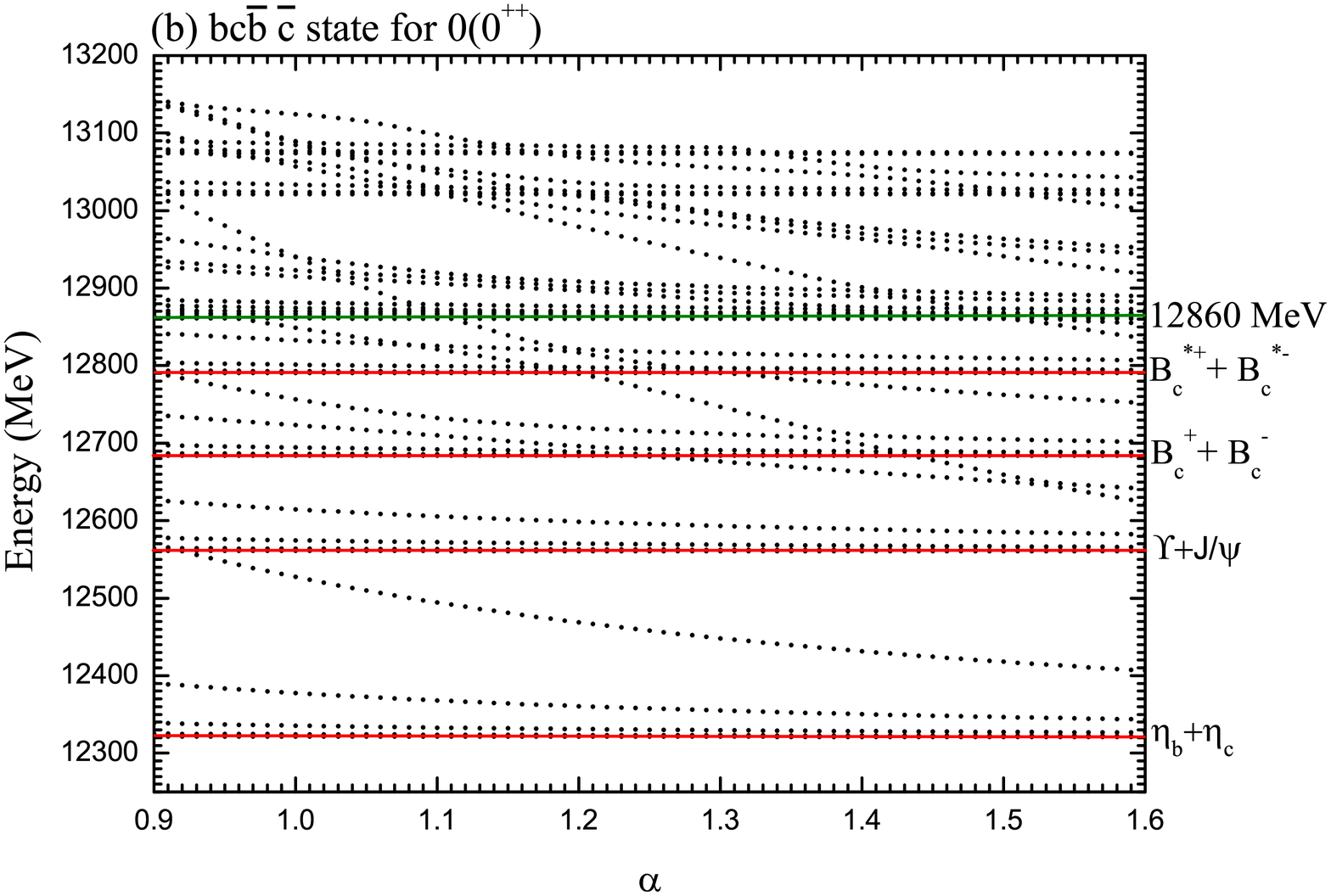}
    \end{minipage}
\caption{\label{rsc00} The stabilization plots of the energies of
$bb\bar{c}\bar{c}$ for $I(J^P)=0(0^+)$ and $bc\bar{b}\bar{c}$
state for $I(J^{PC})=0(0^{++})$ with the respect to the scaling
factor $\alpha$.}
\end{figure*}

\begin{figure*}
\centering
    \begin{minipage}[c]{0.45\textwidth}
    \centering
    \includegraphics[height=6.0cm,width=8.5cm]{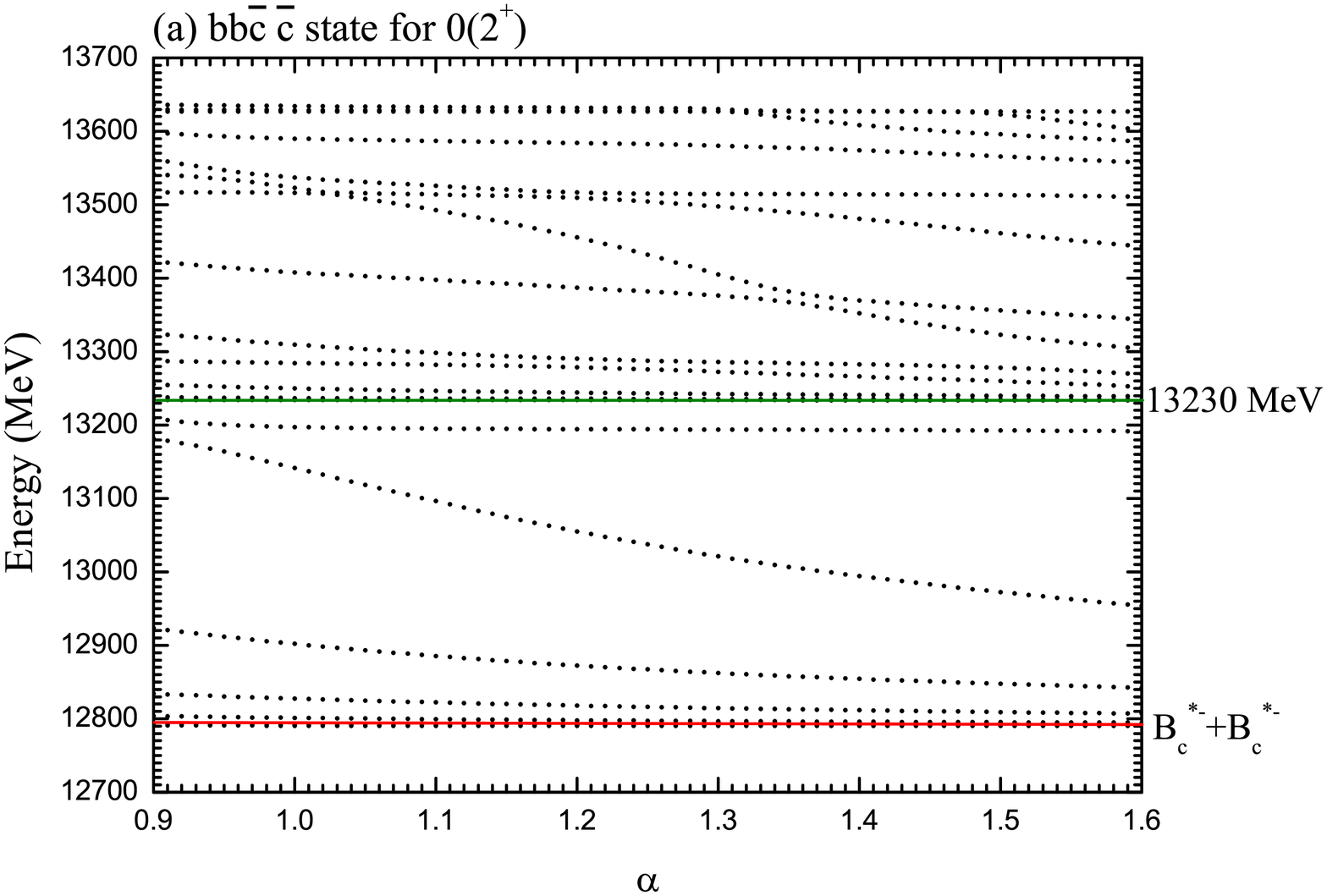}
    \end{minipage}
    \begin{minipage}[c]{0.45\textwidth}
    \centering
    \includegraphics[height=6.0cm,width=8.5cm]{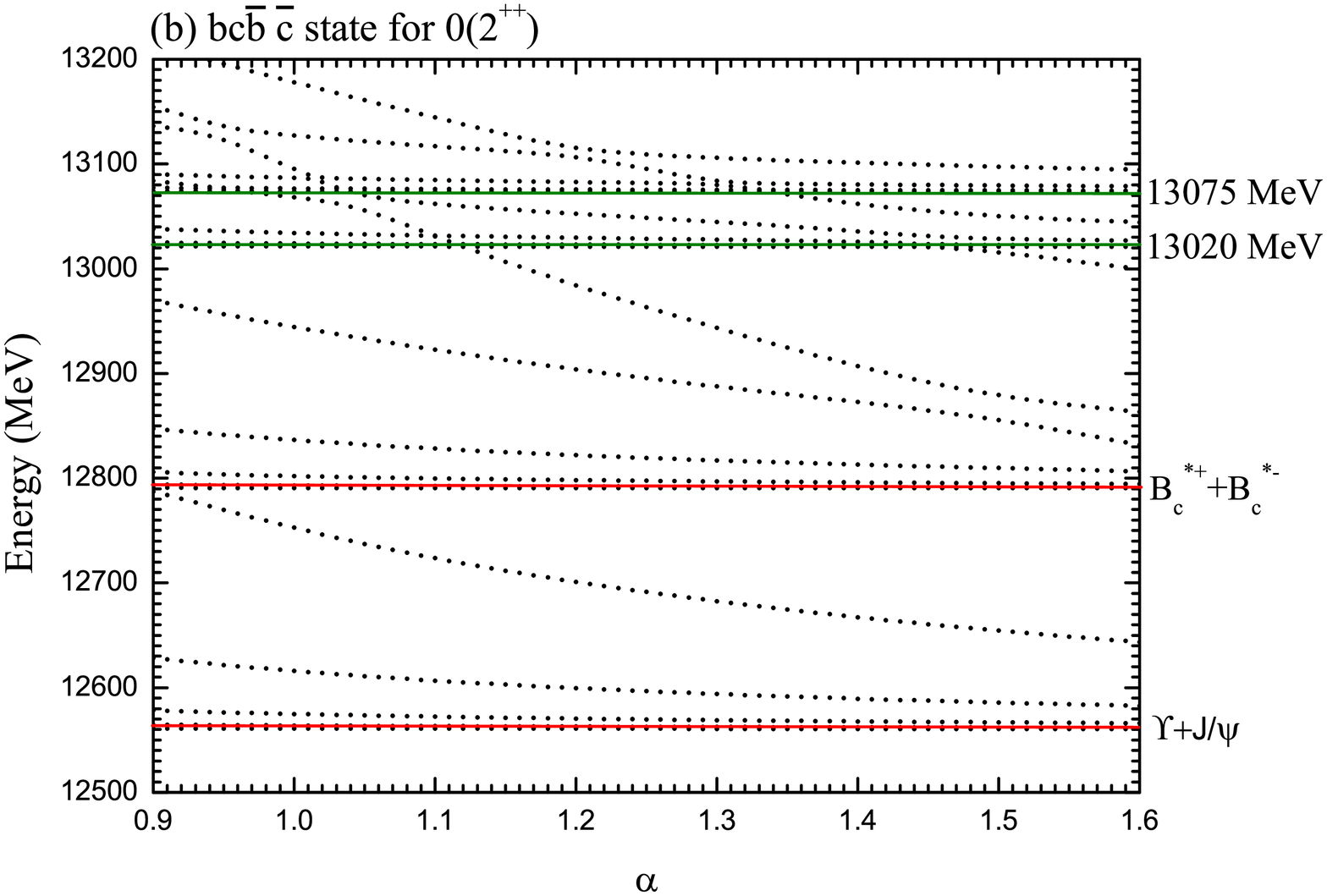}
    \end{minipage}
\caption{\label{rsc02} The stabilization plots of the energies of
$bb\bar{c}\bar{c}$ for $I(J^P)=0(2^+)$ and $bc\bar{b}\bar{c}$
state for $I(J^{PC})=0(2^{++})$ with the respect to the scaling
factor $\alpha$.}
\end{figure*}

\begin{figure*}
\centering
    \begin{minipage}[c]{0.45\textwidth}
    \centering
    \includegraphics[height=6.0cm,width=8.5cm]{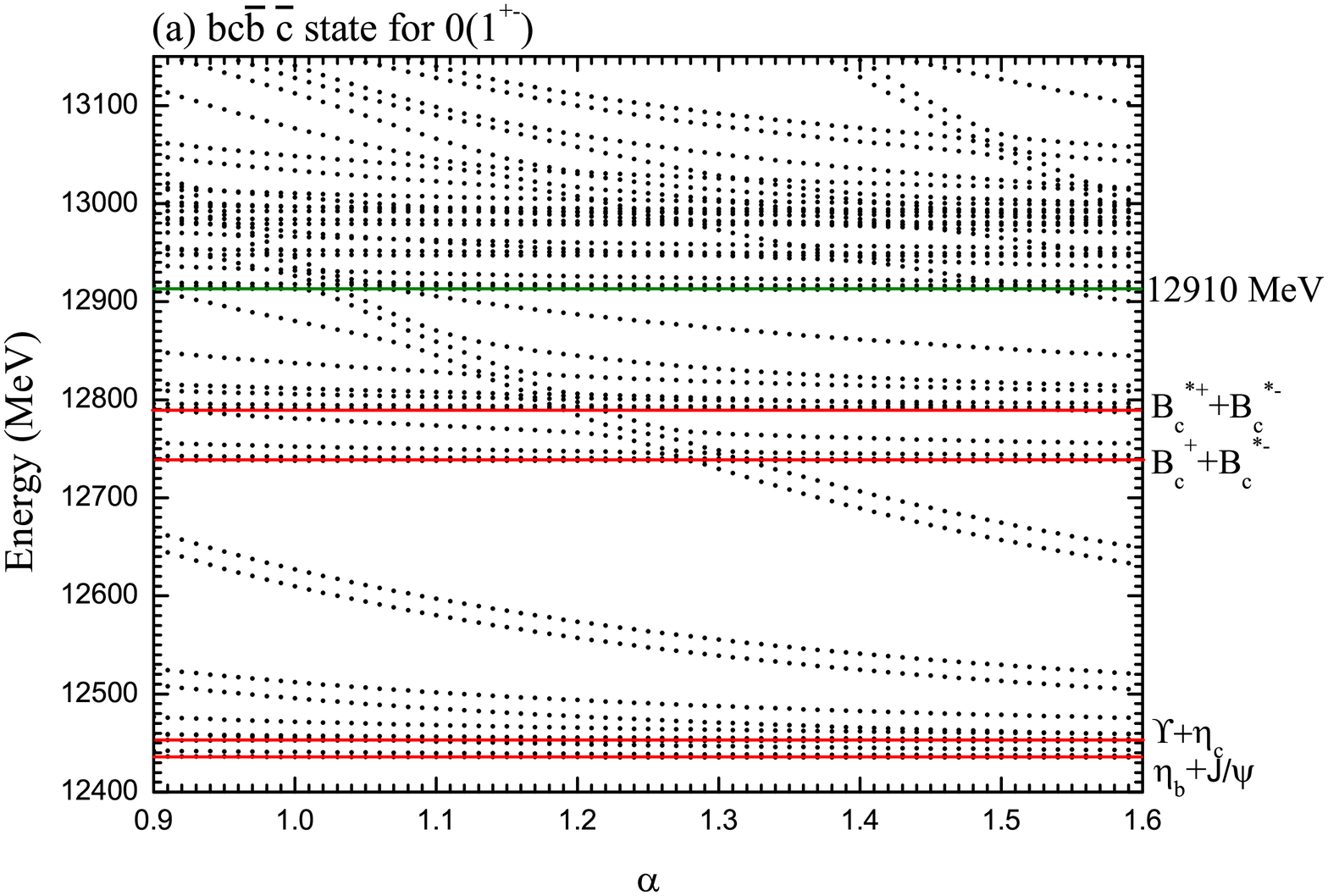}
    \end{minipage}
    \begin{minipage}[c]{0.45\textwidth}
    \centering
    \includegraphics[height=6.0cm,width=8.5cm]{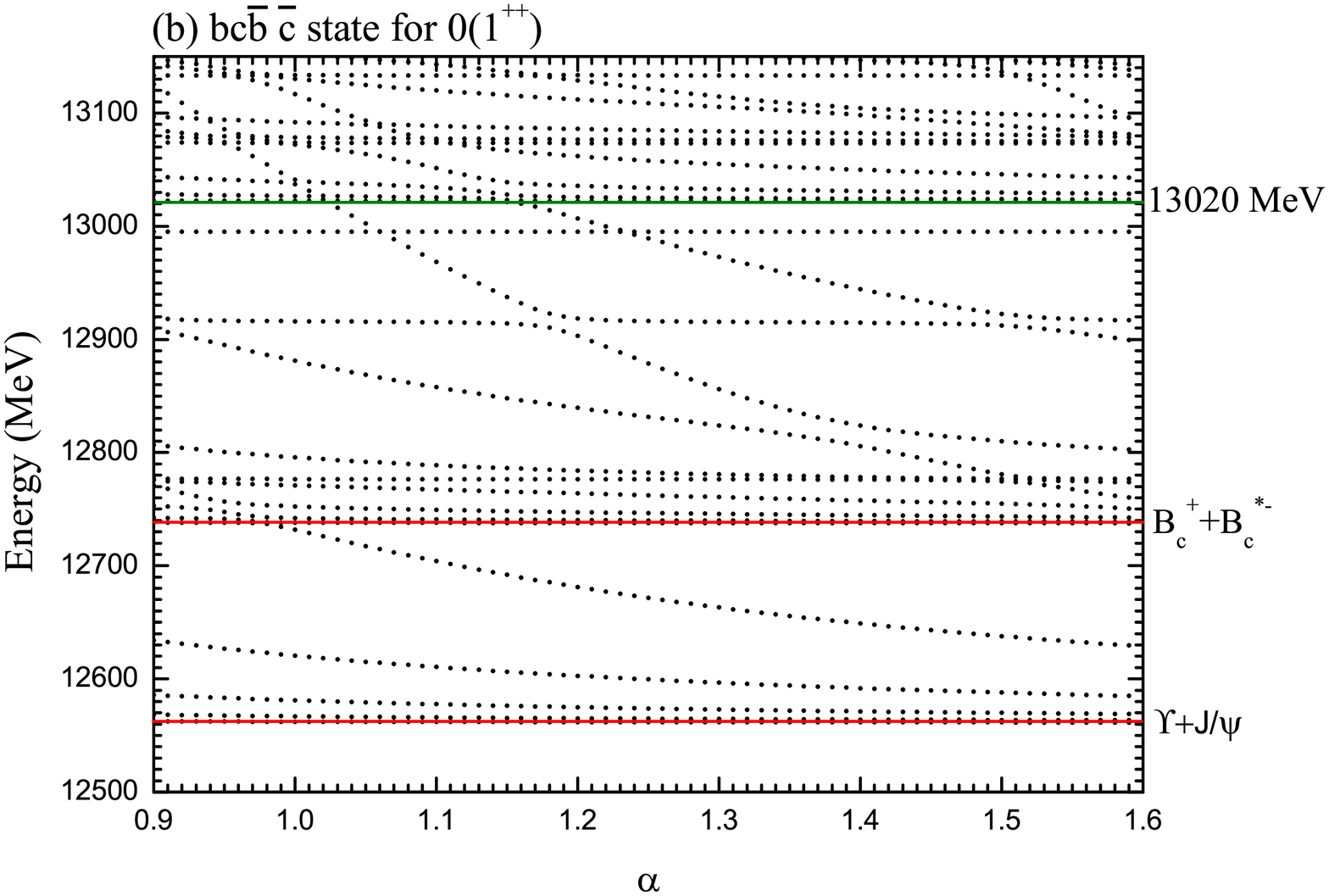}
    \end{minipage}
\caption{\label{rsc01} The stabilization plots of the energies of
$bc\bar{b}\bar{c}$ state for $I(J^{PC})=0(1^{+-}),0(1^{++})$ with
the respect to the scaling factor $\alpha$.}
\end{figure*}

\begin{figure}
\centering
    \includegraphics[height=6.0cm,width=8.5cm]{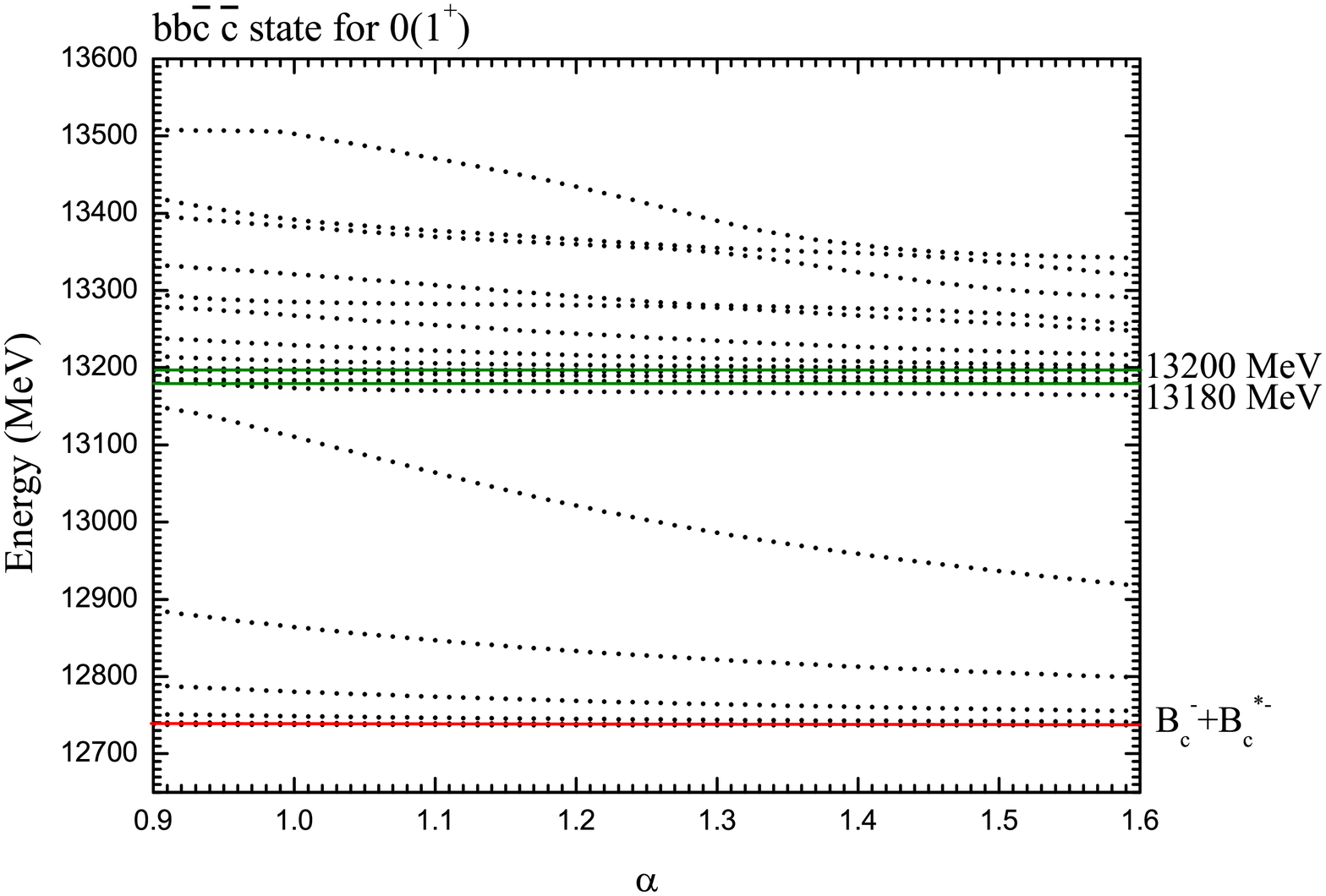}
\caption{\label{rsc01n} The stabilization plots of the energies of
$bb\bar{c}\bar{c}$ state for $I(J^P)=0(1^+)$ with the respect to
the scaling factor $\alpha$.}
\end{figure}

Because the colorful clusters cannot fall apart, there may be a resonance even with the higher
eigenenergy. To find the genuine resonances, the dedicated real scaling (stabilization) method
is employed. To realize the real scaling method in our calculation, the Gaussian size parameters
$r_n$ in Eq.~(\ref{gaussiansize}) are multiplied by a factor $\alpha$, $\alpha r_n$ just for
the meson-meson structure with color singlet-singlet configuration. $\alpha$ takes the values
between 0.9 and 1.6. With the increasing of $\alpha$, all states will fall off towards its
thresholds, but a compact resonance should be stable because it will not be affected by the
boundary at a large distance. We illustrate the results for $bb\bar{c}\bar{c}$ and $bc\bar{b}\bar{c}$
states with all possible quantum numbers Figs.~\ref{rsc00}-\ref{rsc01n}.

In Fig.~\ref{rsc00}, figure (a) represents the $bb\bar{c}\bar{c}$
state for $0(0^+)$. The first two horizontal lines represent the
thresholds of $B_c^- + B_c^- (0\otimes0 \rightarrow 0)$ and
$B_c^{*-} + B_c^{*-} (1\otimes 1 \rightarrow 0)$ for
$bb\bar{c}\bar{c}$ state. The horizontal line around 13140 MeV is
on behalf of a genuine resonance state, and its energy is stable
against the variation of range. For $bc\bar{b}\bar{c}$ state with
quantum number $0(0^{++})$ in figure (b), four thresholds $\eta_b
+\eta_c$, $\Upsilon + J/ \psi$, $B_c^+ + B_c^-$ and $B_c^{*+} +
B_c^{*-}$ are clearly showed, and the first resonance state with
energy about 12860 MeV stays stable.

From Fig.~\ref{rsc02}, we can see that the energy of the lowest
resonance is about 13230 MeV and 13020 MeV for $bb\bar{c}\bar{c}$
and $bc\bar{b}\bar{c}$ state, respectively.

For $bc\bar{b}\bar{c}$ states with $0(1^{++})$ and $0(1^{+-})$ in
Fig.~\ref{rsc01}, the lowest possible resonance is at 12910 MeV
for $C$-parity negative and 13020 MeV for $C$-parity positive. For
$bb\bar{c}\bar{c}$ states with $I(J^P)=0(1^{+})$ in
Fig.~\ref{rsc01n}, two possible resonances stay very close to each
other, with the stable energies 13180 MeV and 13200 MeV,
respectively.

From our calculation, we can see that there may be more resonance states with
the higher energies, and these states may be too wide to be observed or too hard
to be produced. We are interested in the genuine resonance state with as low as
possible energy, so we only give the lowest resonances for each quantum number set.
We hope these information will be helpful for the
searching for the $bb\bar{c}\bar{c}$ and $bc\bar{b}\bar{c}$ states
in experiment.


\section{Summary} \label{epilogue}
In the framework of the chiral quark model, we do a systematical
calculation for the mass spectra of $bb\bar{c}\bar{c}$ and
$bc\bar{b}\bar{c}$ systems with allowed quantum numbers using the
Gaussian expansion method. The meson-meson structure, the
diquark-antidiquark structure and the mixing of them are
investigated severally. In our calculation all these states are
found to have masses above the corresponding two meson decay
thresholds, leaving no space for a bound state. These results are
consistent with our qualitative analysis of properties of the
interactions between quarks. With the help of the real scaling
method, we try to look for the possible resonances in
$bb\bar{c}\bar{c}$ and $bc\bar{b}\bar{c}$ systems. For
$bb\bar{c}\bar{c}$ system, the energies of the possible resonances
are 13140 MeV, 13180 MeV and 13230 MeV for $0(0^+)$, $0(1^+)$ and
$0(2^+)$ state, respectively. For $bc\bar{b}\bar{c}$ system, the
resonance energies are little lower than $bb\bar{c}\bar{c}$ state,
which takes 12860 MeV, 13020 MeV, and 13020 MeV for $0(0^{++})$,
$0(1^{++})$ and $0(2^{++})$ states, and 12910 MeV for $0(1^{+-})$,
respectively. Hopefully, these information about the exotic
tetraquark states composed of four heavy quarks may be useful for
the search in experiments in the future.

\acknowledgments This work is supported partly by the National Natural
Science Foundation of China under Contract Nos. 11847145 and 11775118.



\begin{thebibliography}{10}
\bibitem{x3872}Belle Collaboration (S.-K. Choi \emph{et al.}), Phys. Rev. Lett. {\bf 91}, 262001 (2003).
\bibitem{y4260-1}BaBar Collaboration (Aubert B \emph{et al.}), Phys. Rev. Lett. {\bf 95}, 142001 (2005).
\bibitem{y4260-2}Belle Collaboration (Yuan CZ \emph{et al.}), Phys. Rev. Lett. {\bf 99}, 182004 (2007).
\bibitem{zc3900-1}BESIII Collaboration, (M. Ablikim \emph{et al.}), Phys. Rev. Lett. {\bf 110}, 252001 (2013).
\bibitem{zc3900-2}Belle Collaboration, (Z. Q. Liu \emph{et al.}), Phys. Rev. Lett. {\bf 110}, 252002 (2013).
\bibitem{zc3900-3}T. Xiao, S. Dobbs, A. Tomaradze, and K. K. Seth, Phys. Lett. B {\bf 727}, 366 (2013).
\bibitem{zc3900-4}BESIII Collaboration, (M. Ablikim \emph{et al.}), Phys. Rev. Lett. {\bf 115}, 112003 (2015).
\bibitem{zb10610}Belle Collaboration, (A. Bondar \emph{et al.}), Phys. Rev. Lett. {\bf 108}, 122001 (2012).
\bibitem{pc}LHCb Collaboration, (R. Aaij \emph{et al.}), Phys. Rev. Lett. {\bf 115}, 072001 (2015).
\bibitem{LHCbbbb}LHCb Collaboration, (R. Aaij \emph{et al.}), arXiv:1806,09707 [hep-ex].
\bibitem{plb773247}W. Chen, H. X. Chen, X. Liu, T. G. Steele and S. L. Zhu, Phys. Lett. B {\bf 773}, 247 (2017).
\bibitem{epjc78647}M. N. Anwar, J. Ferretti, F. K. Guo, E. Santopinto and B. S. Zou, Eur. Phys. J. C {\bf 78}, 647 (2018).
\bibitem{prd95034011}M. Karliner, S. Nussinov and J. L. Rosner, Phys. Rev. D {\bf 95}, 034011 (2017).
\bibitem{arxiv161200012}Y. Bai, S. Lu and J. Osborne, arXiv:1612.00012 [hep-ph].
\bibitem{prd86034004}A. V. Berezhnoy, A. V. Luchinsky and A. A. Novoselov, Phys. Rev. D {\bf 86}, 034004 (2012).
\bibitem{arxiv180708520}Z. G. Wang and Z. Y. Di, arXiv:1807.08520 [hep-ph].
\bibitem{arxiv170607553}V. R. Debastiani and F. S. Navarra, arXiv:1706.07553 [hep-ph].
\bibitem{arxiv180706040}A. Esposito and A. D. Polosa, arXiv:1807.06040 [hep-ph].
\bibitem{prd252370}J. P. Ader, J. M. Richard and P. Taxil, Phys. Rev. D {\bf 25}, 2370 (1982).
\bibitem{prd70014009}R. J. Lloyd and J. P. Vary, Phys. Rev. D {\bf 70}, 014009 (2004).
\bibitem{prc97035211}J. M. Richard, A. Valcarce and J. Vijande, Phys. Rev. C {\bf 97}, 035211 (2018).
\bibitem{prd97094015}J.Wu, Y. R. Liu, K. Chen, X. Liu and S. L. Zhu, Phys. Rev. D {\bf 97}, 094015 (2018).
\bibitem{prd97054505}C. Hughes, E. Eichten and C. T. H. Davies, Phys. Rev. D {\bf 97}, 054505 (2018).
\bibitem{epja55106}Xiaoyun Chen, Eur. Phys. J. A  {\bf 55}, 106 (2019).
\bibitem{Li:2019uch} G.~Li, X.~F.~Wang and Y.~Xing, Eur. Phys. J. C {\bf 79}, no. 8, 645
(2019).
\bibitem{Li:2018bkh}G.~Li, X.~F.~Wang and Y.~Xing, Eur. Phys. J. C {\bf 79}, no. 3, 210
(2019).
\bibitem{arxiv190102564}Ming-Sheng Liu, Qi-Fang L\"u, Xian-Hui Zhong, Qiang Zhao, arXiv:1901.02564 [hep-ph].
\bibitem{prd95054019}Jean-Marc Richard, A. Valcarce, and J. Vijande, Phys. Rev. D {\bf 95}, 054019 (2017).
\bibitem{GEM}E. Hiyama, Y. Kino, M. Kamimura, Prog. Part. Nucl. Phys. {\bf 51}, 223 (2003).
\bibitem{plb633237}E. Hiyama, M. Kamimura, A. Hosaka, H. Toki, and M. Yahiro, Phys. Lett. B {\bf 633}, 237 (2006).
\bibitem{prc98045208}E. Hiyama, A. Hosaka, M. Oka, J-M Richard, Phys. Rev. C {\bf 98}, 045208 (2018).
\bibitem{RSM}J. Simon, J. Chem. Phys. {\bf 75}, 2465 (1981).
\bibitem{094016chen} Xiaoyun Chen, J. L. Ping, C. D. Roberts and J. Segovia, Phys. Rev. D {\bf 97}, 094016 (2018).
\bibitem{Vijande:2005}J. Vijande, F Fern\'{a}ndez and A. Valcarce, J. Phys. G. {\bf 31}, 481 (2005).
\bibitem{054022chen}Xiaoyun Chen and J. L. Ping, Phy. Rev. D {\bf 98}, 054022 (2018).
\bibitem{ycyang} Y. C. Yang, C. R. Deng, J. L. Ping and T. Goldman, Phy. Rev. D {\bf 80}, 114023 (2009).
\end{thebibliography}
\end{document}